# Principles of the Battery Data Genome

L. Ward[1], S. Babinec[1,*], E.J. Dufek[2,*], D.A. Howey[3,4*], V. Viswanathan[5,*], M. Aykol[6], D.A.C. Beck[7], B. Blaiszik[1,8], B.R. Chen[2], G. Crabtree[1,9], V. De Angelis[10], P. Dechent[11,12], M. Dubarry[13], E.E. Eggleton[7], D.P. Finegan[14], I. Foster[1], C. Gopal[6], P.K. Herring[6], V.W. Hu[7], N.H. Paulson[1], Y. Preger[10], D.U. Sauer[11,12], K. Smith[14], S.W. Snyder[2], S. Sripad[5], T.R. Tanim[2], L. Teo[7]

[1] Argonne National Laboratory, 9700 S Cass Ave, Lemont, IL, 60439, USA

[2] Idaho National Laboratory, 1955 N. Fremont Ave., Idaho Falls, ID 83415, USA

[3] Department of Engineering Science, University of Oxford, Parks Road, Oxford, OX1 3PJ, United Kingdom

[4] Faraday Institution, Harwell, United Kingdom

[5] Department of Mechanical Engineering, Carnegie Mellon University, Pittsburgh, PA 15213, USA

[6] Toyota Research Institute, 4440 El Camino Real, Los Altos, CA 94022, USA

[7] Department of Chemical Engineering, University of Washington, 105 Benson Hall, Seattle, WA, 98195, USA

[8] University of Chicago, 5730 S Ellis Ave, Chicago, IL 60637, USA

[9] University of Illinois at Chicago, Chicago IL 60607, USA

[10] Sandia National Laboratories, 1515 Eubank Blvd SE, Albuquerque, NM 87185, USA

[11] RWTH Aachen University, Templergraben 55, 52062 Aachen, Germany

[12] Helmholtz Institute Münster (HI MS), IEK-12, Forschungszentrum Jülich, Jägerstrasse 17-19, 52066 Aachen, Germany

[13] Hawaii Natural Energy Institute, University of Hawaii, Honolulu, HI 96822, USA

[14] National Renewable Energy Laboratory, 15013 Denver West Pkwy, Golden, CO 80401, USA

*Corresponding authors: sbabinec@anl.gov, eric.dufek@inl.gov, David.howey@eng.ox.ac.uk, venkatv@andrew.cmu.edu

**Abstract**

Electrochemical energy storage is central to modern society—from consumer electronics to electrified transportation and the power grid. It is no longer just a convenience but a critical enabler of the transition to a resilient, low-carbon economy. The large pluralistic battery research and development community serving these needs has evolved into diverse specialties spanning materials discovery, battery chemistry, design innovation, scale-up, manufacturing and deployment. Despite the maturity and the impact of battery science and technology, the data and software practices among these disparate groups are far behind the state-of-the-art in other fields (e.g. drug discovery), which have enjoyed significant increases in the rate of innovation. Incremental performance gains and lost research productivity, which are the consequences, retard innovation and societal progress. Examples span every field of battery research , from the slow and iterative nature of materials discovery, to the repeated and time-consuming performance testing of cells and the mitigation of degradation and failures. The fundamental issue is that modern data science methods require large amounts of data and the battery community lacks the requisite scalable, standardized data hubs required for immediate use of these approaches. Lack of uniform data practices is a central barrier to the scale problem. In this perspective we identify the data- and software-sharing gaps and propose the unifying principles and tools needed to build a robust community of data hubs, which provide flexible sharing formats to address diverse needs. The Battery Data Genome is offered as a data-centric initiative that will enable the transformative acceleration of battery science and technology, and will ultimately serve as a catalyst to revolutionize our approach to innovation.

**Introduction**

Energy storage is a cornerstone of decarbonization - it enables electrified transportation and increased renewable energy generation on the electricity grid.[1,2] The impact of energy storage continues to expand rapidly as performance steadily improves, costs rapidly decline, and deployments accelerate.[3,4] The Nobel Committee acknowledged the importance of energy storage by awarding the 2019 Chemistry Prize for the development of lithium-ion batteries which "laid the foundation of a wireless, fossil fuel-free society".[5] Energy storage today is an international enterprise with many countries engaging in innovations across the battery value chain. For example, the U.S. Department of Energy's Energy Storage Grand Challenge,[6] the EU's Battery 2030+ research initiative,[7] and the UK's Faraday Battery Challenge,[8] are all focused on organizing a cohesive battery community as the catalyst for an innovative, robust clean-energy economy.

Despite deep commitment and broad excitement, a significant gap which hinders battery innovation is apparent. The revolutionary data science gains enjoyed by other fields have escaped the energy storage community. Battery data and software sharing practices are decades behind contemporary paradigms in other communities such as geonomics, therapeutics, protein crystallography, atomistic simulations, and materials science to name a few.[9–19] (See SI for further detail.) Losses in productivity span the entire supply, production, and use chain, from the sluggish pace of materials discovery, to time-consuming performance and life testing, to the current inability to mitigate degradation and device failures.[20] For example, it took more than 30 years from initial commercialization of Li-ion batteries to achieve today's energy and power density and cycle life.[21,22] In contrast, it took less than a decade for the Human Genome Project (HGP) to transition from a nascent idea to a public release of the complete human genome dataset.[23,24] Analogies between batteries and the various precedents from other fields are imperfect since each has different objectives, for example genomes which unravel a complex existing biological system, but collectively they show that a rich, open dataset with common analysis tools can enable both public knowledge and private wealth and catalyze extraordinary innovation. The global community no longer has the luxury of time in the face of the urgent need to reduce greenhouse gas emissions while supporting broader societal access to energy.

The fundamental issue is that progress in modern data science methods requires large quantities of data, and no single entity can provide data on a large enough scale for these methods to be fully effective. We need community data hubs that accumulate contributions from many groups, but the common practice of keeping battery data in disjointed and incompatible formats that are not easily shared is a central barrier. This failure to adopt a common approach to data organization and standardization with a common language is a critical gap that has prevented the majority of available data from being used for machine learning (ML). The solution to this scaling problem is to improve uniformity across battery data practices. The Battery Data Genome (BDG) aims to provide a path to rich data resources by establishing uniformity in standards, software and tools and by creation of a variety of data hubs that can serve diverse needs across the entire ecosystem. In the near term, data-driven approaches such as ML and artificial intelligence (AI), can dramatically improve efficiency in research and development as well as expand the scope of problems addressed. Ultimately these new capabilities will revolutionize our approach for energy storage innovation and enable methods such as inverse design (i.e. directly identify materials that will provide desired functionalities) and direct extrapolation from material properties to real-world performance.

We assert that a rapid catch-up is possible by connecting data-science best practices with the extensive data stockpiles and the deep scientific capabilities of the battery community. Conceptual proofs are already available for the ubiquitous and time-intensive battery life prediction step. Recent studies used ML-methods to predict cycle life from as little as 100 initial cycles,[25,26] infer degradation paths using a synthetic dataset,[27] and identify early indication of Li plating, a major safety concern, by analyzing cycle-by-cycle electrochemical data.[28] Discovery of new battery materials with targeted properties has been significantly accelerated by inverse design using ML[29,30] and fully autonomous materials discovery laboratories are now emerging.[31–35] Despite this early promising progress, such success stories are infrequent due in large part to limited availability of shared data in coherent and accessible formats.

Modern data science best practices combined with lessons from previous big-data projects set a rational expectation for what is possible, a framework for action, and a vision that is both aspirational and practical. As with other data initiatives that address complex systems,[11,36] the objective here is to accelerate the pace of discovery at a fraction of the current cost in order to provide transformative impacts. In this perspective, we propose that the community launch the Battery Data Genome as an initiative which creates a collection of data hubs, whose standards and practices are completely open, along with flexible sharing practices, with contributions from research, manufacturing, and deployment, across all battery chemistries. A foundational premise is that a portion of the battery community will not or cannot provide open access but can still leverage the standardized formats and flexible sharing options, and this will enable the broad participation needed for impact. We suggest that the massively complex and heterogeneous energy storage data be organized based on how the data is generated, which characteristically aligns with technological stage of development, and offer a participation hierarchy to address disparate stakeholder objectives. A common language is an essential means to this end - it consists of both standardized data formats and shared open software which can accommodate standardized interoperable functions. Together these set the stage for knowledge integration across domains that seeds innovation breakthroughs.

**Heterogeneity and Scale Challenges**

The BDG faces two significant challenges: establishing the data and metadata conventions that will make heterogeneous data useful and enable interoperability, and rapid, large-scale capture of data from many sources and contributors. Heterogeneity is created by multiple phenomena covering a range of length scales, from molecular to pack dimensions, and time domains from seconds to decades - see Figure 1. Rapidly evolving experimental and developmental capabilities allow researchers to generate diverse data across these spectrums at unprecedented scales and precision. Heterogeneity is further compounded by bench scale reproducibility, cell-to-cell manufacturing uniformity, and thermal control.

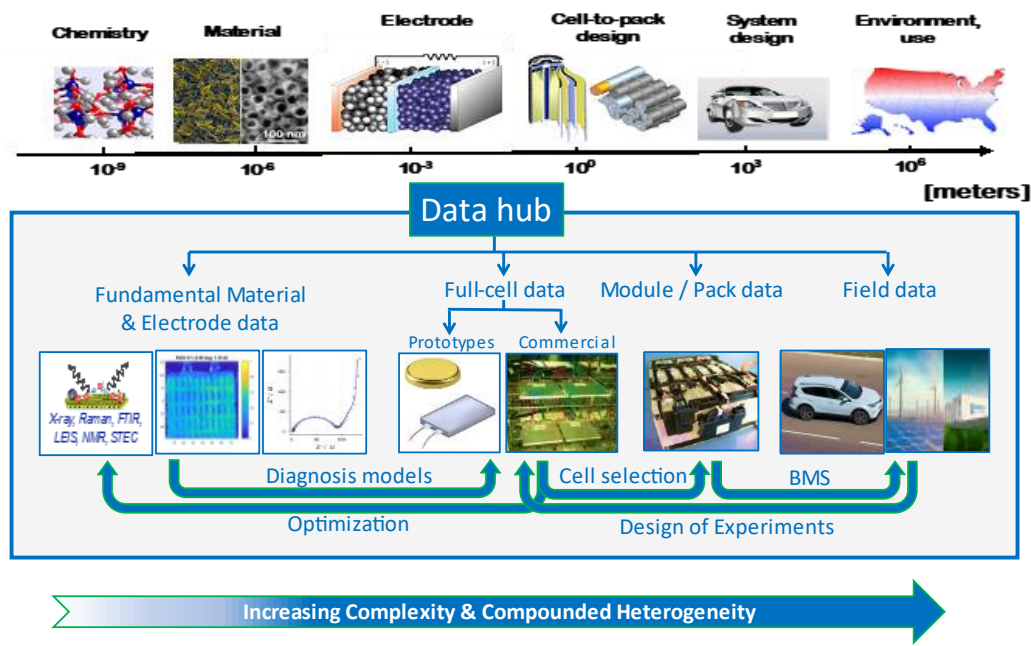

Figure 1. The four primary segments of the Battery Data Genome

**Organizing Principles**

Both heterogeneity and scale challenges are addressed by organizing the data into four segments based on the functionality of the battery component or system under study. Division into four segments allows data associated with each segment to share a common format; each segment can then accommodate large quantities of data from various measurement sources while maintaining interoperability.

The four segments are described below and illustrated in Fig 1.

1. **Fundamental material and electrode studies** are used to probe bulk and interfacial processes at the atomic and molecular scale[19,37–40], for both materials discovery and commercial electrodes.[41–43] For example, materials characterization can include electrochemical data and spectroscopy, x-ray characterization, etc. Electrode characterizations can include optical, acoustic, post-test tear down or synchrotron-based studies.

2. **Full cell evaluations** reveal the interplay among materials, cell designs, and performance for

prototype[44,45] and commercial single cells.[26,46–48] For different battery designs such as flow cells the nature of the data may differ from a Li-ion or lead acid cell. This level of data can also include non-electrochemical data.

3. **Module/pack characterization** reflects system performance in a controlled environment[49] and the performance of a design in managing cell-to-cell heterogeneity – both with respect to initial variability and as the system ages.

4. **Field data** provides insights into real-life performance, providing valuable understanding of gaps between lab and field, and the research needed to fill those gaps.

This data segmenting approach is intended to meet the needs of a wide range of chemistries, with the principles that apply to system-specific requirements. For example, while Li-ion technology is presently an active field, the organizing principles will translate to Li-metal, new low-cost aqueous chemistries, flow cells, and the many evolving chemistries for longer duration storage.

Early examples of battery data archives fit these organizing principles. The Battery Archive was created to provide data for battery degradation studies;[48,50] the Battery Evaluation and Early Prediction (BEEP) tools are initially focused on optimization of fast-charging protocols for batteries;[51] and Galvanalyser aims to unify the process of gathering and querying data from different types of battery testers. Emerging examples aligning with the fundamental material and electrode segments include several significant databases which link battery chemistry to material properties[53] and electrode microstructure to performance.[37,54–57] There are also databases that link battery information to other types of information; e.g., various EV battery lifetime tools[58] leverage open meteorological data[59] and drive cycles[60]. These specialized data banks can serve as models and guides to establishing the structure and metadata requirements for BDG.

**Operating Principles**

Four principles describe essential operations across all four data segments and enable the BDG to function predictably and thereby insure the broad participation required for maximum impact.

*The first operating principle is that uniform standards and protocols guide how experiments should be performed, how existing data can be adapted, what types of data should be collected, and ultimately how we train researchers. Uniformity is essential for interoperability which is critical for breakthrough innovations arising from cross-fertilization.* [61–69] The types of data fall broadly into (i) materials characterization data such as x-ray diffraction, transmission electron microscopy, x-ray, infrared and Raman spectroscopy; (ii) experimental cell or pack performance data, which is often time series measurements of current, voltage, temperature, or electrochemical impedance spectroscopy data, (iii) synthetic data spanning materials to packs, i.e. generated by models, and (iv) 'data about data' (metadata), i.e., information about the cell and system creation, experimental setup, and so on.

*The second operating principle is that metadata is as important as performance data, is complex and heterogeneous, and requires detailed reporting protocols to insure adequacy and consistency.* In addition to electrochemical information, metadata should include information on the chemical, structural, and physical characteristics of the cells and materials involved in an experiment. It is noted that characterization information only belongs in metadata when a well-established reference exists that correlates with properties; if the significance of much of the data is not understood or established, then this information is experiment data, not metadata. In both cases, this information is important for the BDG.

Standardized metadata across a broad interdisciplinary community is required to extract the maximum value and impact. Establishing what information is required and the format in which it should be recorded and stored is a major challenge that requires extensive consideration and innovation. For example, in battery material synthesis and manufacturing studies, it can be very challenging to find a consistent approach for the extensive documentation needed to ensure value and impact. Full cell and electrode metadata should include information on testing conditions, references and protocols used, additional characterizations (such as safety), and controls or experimental details, all reported in standardized ways. Additional detail on metadata is in the SI.

Since broadest possible community participation is an objective and since full disclosure is not an option for certain stakeholders, the BDG will accommodate various needs by pragmatically organizing metadata into *primary*, *secondary*, and *tertiary* layers of detail (Figure 2a).

Primary metadata is a small set of mandatory items that must be reported to provide the basic concepts of the data for easy searchability. It includes *minimal high-level identifying information*, and can be viewed as the table of contents for each specific data set. It should include the owners and generators of the data, a unique identifier for each cell, the objective of the experiment (for example electrochemical characterization, or materials discovery), the nature of the device/s that were tested including chemistry, and specific tests. This level is rigidly structured to ensure consistency and interoperability across data sets.

Secondary metadata (Figure 2) is highly detailed and requires a structure that ensures searchability and interoperability. *Critically, secondary metadata triggers scientific insights by enabling users to map between the design variables, test conditions and output performance metrics.* Key aspects of secondary metadata include further details of the batteries under test such as cell and electrode design information, commercial cell identification codes, the manner in which cells were attached to test equipment, and additional test details. Many of these could follow emerging reporting requirements, e.g. the checklist created by the journal Joule.[70] Specific reporting requirements for materials discovery need to be defined by the appropriate stakeholders. The second tier contains significant experimental details, and thus has immense value for discovery and mechanistic insights. Researchers should aim to provide as much detail as possible in this second tier, keeping in mind the balance between shared and proprietary information.

Tertiary metadata may be required when specific sub-groups within the community are interested in specialized information. For example, in materials discovery, synthesis protocols have a significant impact on performance - tracking the specific steps taken during synthesis will be vital to the materials community but is less broadly relevant when using mass produced or commercial materials. The structure and content of tertiary metadata could vary substantially from sub-field to sub-field and needs to be defined by consensus among the stakeholders of the subfield in question.

Figure 2:

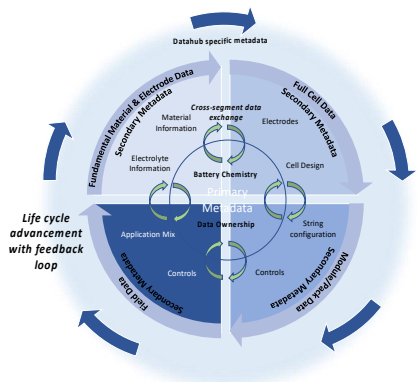 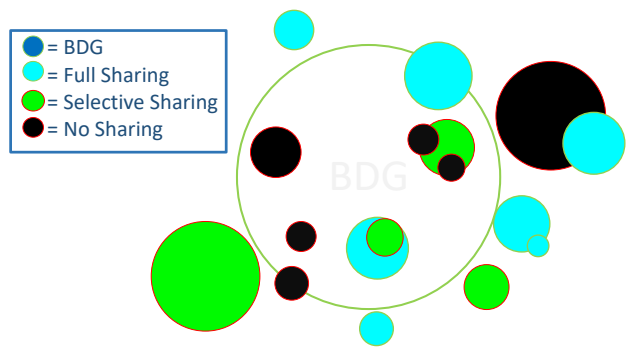

FIGURE 2A: *Organizing principles:* Each of the four proposed segments has a hierarchy of primary, secondary and tertiary metadata

FIGURE 2B: *Sharing principles:* Data hubs, represented as circles, all operate with the same organizing and operating principles, but not all openly share data. Here, extent of overlap represent extent of sharing. The large and fully open BDG datahub will contain all information segments. Other hubs will not necessarily have all segments.

*The third operating principle is that all performance and metadata must be cleaned, curated, and controlled before it can be used for downstream purposes.* Relevant data covers everything from lab materials and cell scale characterization measurements with various use-protocols, to actual field deployment information collected over the lifetime of the system. Battery characterizations, whether related to performance, degradation, safety, or materials characterization, typically comprise current/voltage/time profiles, and must allow for more sophisticated future characterizations. Test protocols vary from bespoke procedures used for early research up to rigorous evaluation standards for commercial uses.[71,72] There are at least two

significant complications in converting measured performance data into efficient and easy to query formats: (1) output current/voltage/time/temperature data often requires clean-up with uniform agreed upon standardization, quality metrics, and strategies to resolve gaps; (2) the sheer volume of data suggests the need for efficient storage and back-up. The solution to these issues can be driven by collaboration between working groups who understand both the electrochemical science and the data science paradigms.

*The fourth operating principle is that not all data need be openly shared*. The BDG concept is for one central, identified data hub with the purpose of maximum open sharing, and for unlimited additional data hubs with many levels of engagement, partnerships and sharing – See Figure 2b. We anticipate that the one very large data hub will have many contributers and that the data will be structured into the four segments of Figure 2a. We further anticipate that additional data hubs would be beneficial for different scientific and commercial objectives and may not need to include all four segments– such as materials discovery vs. performance validation vs.manufacturing excellence. Our assumption is that the scale of data necessary for breakthrough innovations is far less than the total amount of data which exists, and thus success does not require full community disclosure. *To be effective, however, all must operate with* the same operating principles and the common framework.

**Software Principles**

Fragmented software development has hampered data-science advances and exacerbated reproducibility issues in batteries. Advanced software is a critical supporting function of the BDG and it would be a missed opportunity if we did not also reimagine its role. Shared software tools that automate data management and testing, and draw new understanding from data are needed to accelerate innovation.

We envision *at least one complete open-sourced suite* of interoperable software for translating data from heterogeneous sources into the shared format of the BDG and standardizing experimental protocols. This enables coordination across standardized data hubs and comparison of data against testing standards, so as to provide the common approach required to create a community. It also supports the streamlining and uniformity of test procedures and hardware configurations across all stages of development and types of analysis.

Machine learning and AI tools for batteries are currently in their infancy. For example, despite advances in molecular ML, there has been limited success in designing new battery materials. Similarly, cycle life prediction using a physics-based cell degradation model or a data-driven battery health model on small sets of lab data are at a nascent state. Across these disparate topics, extraction and identification of useful features directly from the data and meta-data requires significant improvements including software and data methods which deal with uncertainty (e.g. measurement error, model/inference error), variability (e.g. materials composition, manufacturing) and varied operating conditions.

A key principle is that experimental data can be supplemented with synthetic data. For example, the use of physically-rooted electrochemical performance or aging models provides the opportunity to run virtual experiments ("digital twin") that predict the outcomes of costly and time-consuming physical experiments. This can "close the design loop" between simulation and experiment by uniformly processing and analyzing multi-fidelity datasets. Synthetic data also provides opportunity to benchmark the predictive performance of ML algorithms. The ability to coordinate experimental and synthetic data access through the BDG could provide the seed for partially autonomous laboratories.

Early effort towards community-driven software codes has contributed to solving some of these challenges of data processing and analysis[51,52,73–77] including simulation frameworks (e.g., PyBAMM[73]). Remaining significant software gaps include: (i) Battery cycling protocols that are not standardized or easy to translate between different experimental equipment; (ii) Workflows for data management – parsing, validation,

sharing – are idiosyncratic and data formats are highly non-uniform; (iii) AI/ML code for analysis and automation is a new area, with embryonic code that is ad hoc, lacking modularization and inter-operability. Further details on these gaps are provided in the SI.

The realization of BDG software principles requires fostering linkages among existing software projects and establishing an interconnected community of researchers. To that end, we have developed a proof-of-concept integration between Battery Archive[50], BEEP[51] and PyBAMM[73] (available at Github[78]). These three codes represent complementary capabilities within the battery data ecosystem, with Battery Archive providing long-term archival storage and interactive visualization of cycling data and benefits from data parsers available in BEEP, and PyBaMM offering a framework for simulating the cycling protocols *in silico*. Identifying similar relationships between codes and performing similar integrations is both a community organization and technical challenge.

**Challenge problems drive community engagement and address broad issues**

Challenge problems play a critical role in structuring AI research and development and accelerating its impact. "ImageNet" is a prominent example where a database of about 14 million annotated images, combined with the "ImageNet Large Scale Visual Recognition Challenge," spurred groundbreaking image-classification innovations,[79] developed a large talent ecosystem, and ultimately helped to accelerate the development of vision systems in autonomous vehicles.[80] We envision that BDG challenge problems will similarly catalyze a broad range of new capabilities and unleash otherwise unavailable possibilities as available data scales up – for instance, by enabling autonomous "self-driving" system optimization of battery material/device design, or by determining the source of variability in commercial cell performance. In the spirit of the ImageNet Challenge, we outline five BDG challenge themes for quantitative benchmarking of state-of-the-art methods, framed to ensure their relevance beyond any specific chemistry.

Each area encompasses multiple challenge problems, for example designing electrode materials through AI-accelerated cycle life testing. These challenges will benefit the battery community by (i) enabling assessments of progress in developing predictive algorithms and models,[81] and (ii) providing the foundation for eventual deployment of these algorithms into a full complement of energy storage systems.

While the BDG is organized around characterizing the functionality of the component or system, the challenge problems align with how data is *used* and provide the opportunity to innovate across different segments and methods of data generation. They are the architypes of cross fertilization.

1. **Components**: Given the constituent materials, composition, structure, defects, and manufacturing process for a component (e.g., electrode, electrolyte), predict its performance characteristics. The component performance will serve as an important feature for predicting the overall device performance characteristics. In the realm of materials discovery, an example challenge could be to provide cathode candidates with sufficiently low volume expansion to enable long cycle life in a cell with an inorganic, solid electrolyte.

2. **Interfaces**: Given the materials on both sides of an interface (with component data as listed in #1), predict the chemical composition and structure of the interface and its performance for selective transmission of working ions, blocking of electrons and unwanted complexes, and degradation routes.

3. **Cells**: Given cell architecture, components (described as in #1 and #2), interfaces and cell fabrication rules, predict the performance, life, and safety of the cell under a range of use conditions. We further outline five related subareas that are presented in detail in the SI: a) Performance, b) State of Health, c) Life, d) Safety, and e) Manufacturing.

4. **Packs to Systems**: Given cell materials and components, cell design variables and manufacturing

variability, cell topology, and pack design/control choices, predict pack performance, life, and safety under application scenarios including transportation, grid storage and aviation under variable loads.

5. **Inverse Design**: Invert the challenge problems above – given a set of expected performance characteristics from the end-use application, identify optimal (1) materials, (2) components, (3) interfaces, (4) cell designs and (5) pack designs and controls.

Until suitable datasets are available for the significantly difficult challenge problems, we suggest a two-tiered approach that balances the need for rapid start driving community engagement and an expanding scope with greater complexity. Tier 1 challenges will use presently available data, and more complex Tier 2 challenges will emerge as appropriate open data sets become available.

We offer cycle life estimation as a Tier 1 challenge problem that leverages available fast charging data[26] supplemented by synthetic data[27]. A Tier 2 version would leverage standardization and designs of experiments[82,83] to explore cycle protocol variability as a function of C-rates, average state-of-charge, and temperatures, and to invoke complicated path-dependent degradation. Investigating these issues, and exploring the impact of variability in degradation and performance, will require a significantly larger dataset than is publicly available. To establish a sufficient data hub, we propose that the community launches a crowd-sourcing strategy to reduce the experimental burden on individual researchers[84]. Given the wide variation in degradation modes, large synthetic datasets[27] that cover a broader range of conditions than are practically experimentally accessible, can complement experimental data in both Tier 1 and Tier 2 challenges, and provide a useful comparison. Industry may make data available if it clearly incentivizes innovation and assists in recruitment. As one example, Western Power Distribution (UK) provides their open-source energy data as a catalyst for design of an optimal schedule for battery storage.[85]

The SI offers an expanded Tier 2 experimental design that addresses numerous challenge problems in the

'cell' data segment. This challenge will optimally require data from several thousand cells, and thus may be best addressed by using "crowd-sourced" participants who submit proposals to produce data that meet community experimental design and standard practices and to conduct experiments with preconditioned cells stored in a central community User Facility. Funding to provide, condition, and house thousands of cells at the User Facility would be substantial and would need to be supported by multiple funding agencies.

The present Tier 1 and Tier 2 challenges are dominated by existing Li-ion chemistries as these offersignificant data, but there is a distinct need for future Tier 2 challenges to include emerging battery chemistries and designs. Possible areas for future releases include flow-battery electrolyte systems, materials discovery, other alkali metal-based chemistries, and multi-valent systems. An example is for an electrode satisfying multiple performance criteria, such as high or low operating voltage, high working ion capacity and mobility, low expansion on intercalation, and minimal side reactions. The challenge is to satisfy all the requirements simultaneously while negotiating the tradeoffs.

**Battery Data Genome: Urgent Need and a Path Forward**

Slow progress in the face of many complex science and technology options combined with the desire for a resilient, low carbon economy is driving an urgent need for transformational energy storage technology. Complexity and competition have fractured the pluralistic battery community, which is progressing primarily through incremental advances and which has thus far failed to pursue the evident benefits of contemporary data science practices. The fundamental issue is that progress in data science requires a significant amount of appropriate and standardized data, and no single broadly accessible source is able to provide data at such a scale. Further, battery data is kept in disjointed and incompatible formats which create a barrier to ready coalescence due to a lack of common practices. The solution to this problem of scale is to drive a minimum level of uniformity across battery data practices. Our vision is that the BDG will catalyze innovation that broadly benefits public and private stakeholders via standardized and curated data, seamless data exchange, and interoperable software. The range, scale, and complexity of problems that can be addressed will be dramatically expanded and the cross-fertilization needed for breakthroughs will be made possible. *The remedies for these barriers requires broad adoption* of uniform practices across boundaries between the data science and battery communities and among academic, industrial and startup communities that span the battery ecosystem. Launching this process will require deliberate and intentional cooperation among stakeholders from materials discovery to battery cell design, manufacturing, pack integration and deployment.

Bringing the diverse battery community together is challenging. Competition for visibility, research funding, profit, and market share serve as a strong counterargument to cooperation. At the root of the BDG are stakeholders – individual contributors, public institutions, private profit-generating organizations, and a broad customer base. The stakeholders who will collect and oversee data and software resources and serve as advocates for community engagement must balance their individual aspirations with essential sharing requirements of this vision. Public institutions generally align with broad BDG goals, provided that success metrics, such as publications, are not compromised. Profit-generating organizations will benefit from the use of open community data and software while protecting their trade secrets and propriety information with flexible sharing practices. *We believe that the promise of transformative capabilities that will revolutionize the energy storage industry will sufficiently motivate broad adoption of BDG-standardized practices and interoperable tools and will foster cooperation within a respectfully cohesive community.*

Modern ML and AI can be effective at driving disruptive innovation, but there needs to be data of a sufficient scale and diversity to enable the change. To accommodate the requisite abundant data, multiple data hubs will need to be united by common practices and standardization. A common framework is urgently needed. While we expect that the BDG will only capture a portion of the global battery data, this should be sufficient to drive the broad adoption of the common framework and best practices. To estimate the requirements for success, data-related questions such as "how much data is enough?" must be asked even though they will be difficult to answer in many cases. Notably, participation in the HGP was not uniform and extensive private data genomic data remains to this day. Examples of suitable tracking metrics are offered in the SI.

Launching the BDG will require a multi-phased approach (Figure 3), which could be facilitated by two types of volunteer committees representing the stakeholder community. An implementation group will ultimately be the driving force for BDG execution, while a more expansive stakeholder steering committee will ensure that public and private stakeholder needs are considered and accomodated. Committee memberships should rotate, similar to boards of professional societies. Coordination across all domains and stakeholders will be vital to establishing a standardization and interoperability framework through a series of broad community workshops.

In Phase 1, the focus will be on creation of a comprehensive structure and a diverse yet cohesive community through workshops. More details on workshops are in the SI including those focused on setting standards, a strategy for interoperability, and establishing a roadmap of increasingly demanding targets to achieve the BDG goals. In Phase 2, data hubs are populated with new and published data and become interoperable with software; several challenge problems are pursued, and roadmaps are refined in light of the community's ability for cross-fertilization. The focus of Phase 3 is broadening acceptance and transitioning into a sustainable mode via discussions at professional meetings and focused workshops.

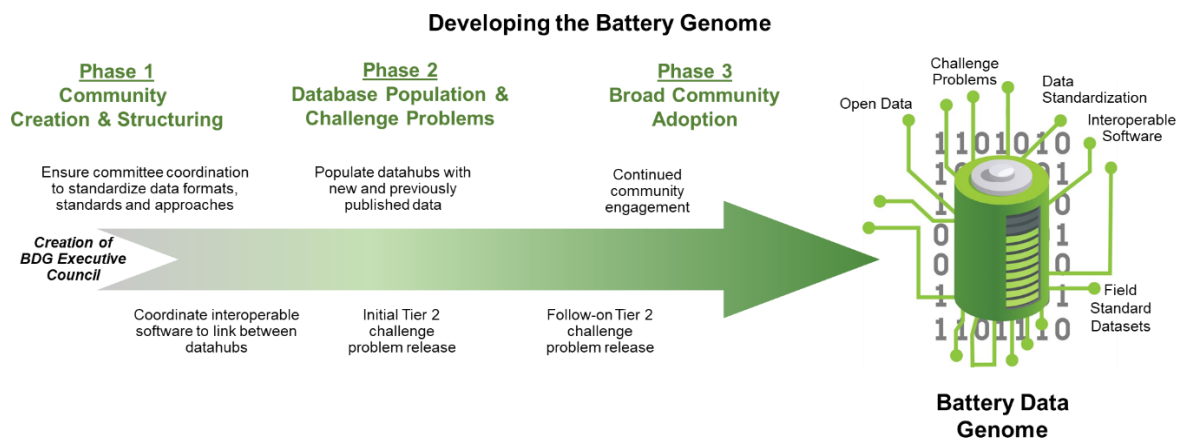

**Figure 3:** Implementation roadmap for the Battery Data Genome

**Concluding Remarks**

We propose the launch of a large-scale energy storage initiative for the adoption of contemporary data science practices in research, development and deployments and outline the groundwork needed for a universal set of data- and software-sharing practices. This will enable more immediate and widespread application of ML and AI in batteries. The fundamental issue is the need for large amounts of common data, the barrier is lack of uniform practices, and the solution is to build a robust community of multiple data hubs with shared core principles. There will be one large, central, open data hub that integrates with other data hubs. We contend that the resulting capabilities will not simply accelerate innovations, but ultimately revolutionize how we approach energy storage innovation by enabling opportunies such as inverse design and direct extrapolation of material properties to performance in the field. Our guiding assumption is that using modern best practices for data science and data collection combined with lessons from previous big-data projects sets a realistic expectation and provides a framework for action. Our organizing principle is to categorize information based on characterizing the functionality of the component or system; our operating principle is for a uniform language and standardized practices, with flexible options for sharing of curated data. Open-sourced interoperable software with common routines and challenge problems further drive broad community engagement. In this perspective we reason that while the goal of coordinating the global battery community and launching the Battery Data Genome may seem grandiose – ultimately it will be found both plausible and a necessary step in combatting climate change and providing energy security.

# Figure captions

**Figure 1:** Different subfields use different data for different purposes, ranging from half-cells exploring material-scale mechanisms to field data capturing the effect of regional environments. A range of BDG Data hubs and challenge problems are therefore required to advance the use of open data in battery science.

**Figure 2a:** *Organizing principles:* Each of the four proposed segments has a hierarchy of primary, secondary and tertiary metadata

**Figure 2b:** *Sharing principles:* Data hubs, represented as circles, all operate with the same organizing and operating principles, but not all openly share data. Here, extent of overlap represtns extent of sharing. The large and fully open BDG data hub will contain all information segments. Other hubs will not necessarily have all segments.

**Figure 3:** Implementation roadmap for the Battery Data Genome

# Author contributions

The ideas described in this paper were refined over the course of a year of video conferences. L.W. organized the meetings and subsequently the writing of this perspective article, with guidance from V.V. and S.B.  S.B., and E.D. led the editing of the final draft, with significant input from S.S., D.H., V.V., G.C., and I.F. The technical groups were led by E.E. (Organizational and Operating Principles), C.G. (Software Principles), V.V. and N.P. (Challenge Problems), and E.D. (Urgent Need and Path Forward). All authors contributed to the concept development and writing of this manuscript.

# Acknowledgements
The authors would like to recognize the contributions from each of the reviewers. The recommendations and items brought forth for consideration aided considerably in the preparation and ultimate impact of this manuscript. S.B, J.K., N.H.P., and L.W. acknowledge financial support from Laboratory Directed Research and Development (LDRD) funding from Argonne National Laboratory, provided by the Director, Office of Science, of the U.S. Department of Energy (DOE) under Contract No. DE-AC02-06CH11357. M.D. acknowledge support from the Office of Naval Research # N00014-18-1-2127 and from the State of Hawaii. Y.P. and V.D.A. acknowledge support from the U.S. DOE Office of Electricity, Energy Storage Program and Sandia National Laboratories. Sandia National Laboratories is a multi-mission laboratory managed and operated by National Technology and Engineering Solutions of Sandia, LLC., a wholly owned subsidiary of Honeywell International, Inc., for the U.S. DOE's National Nuclear Security Administration under contract DE-NA-0003525. K.S., E.D, T.T, B.C and S.S acknowledge support from the U.S. DOE Vehicle Technologies Office. The National Renewable Energy Laboratory is operated by Alliance for Sustainable Energy, LLC, for the U.S. DOE under contract No. DE-AC36-08GO28308. Idaho National Laboratory is operated by Battelle Energy Alliance under Contract Nos. DE-AC07-05ID14517 for the U.S. DOE. The views expressed in the article do not necessarily represent the views of the DOE or the U.S. Government. The U.S. Government retains and the publisher, by accepting the article for publication, acknowledges that the U.S. Government retains a nonexclusive, paid-up, irrevocable, worldwide license to publish or reproduce the published form of this work, or allow others to do so, for U.S. Government purposes. P.D. and D.U.S. gratefully acknowledge the financial support by the German Federal Ministry of Education and Research (BMBF) for funding within the research cluster greenBattNutzung (03XP0302C) and the German Council of Science and Humanities for funding of the Center for Ageing, Reliability and Lifetime Prediction of Electrochemical and Power Electronic Systems (CARL). E.E.E, V.W.H. and L.T. acknowledge Professor Daniel T. Schwartz at the University of Washington and support provided by Army Research Office contracts #W911NF1710550 and #W911NF1810171 for developing electrochemical data science activities in partnership with the Electrochemical Society. D.H. acknowledges funding from Research England's Connecting Capability Fund (`Pitch-In' project, CCF18-7157) and from the Faraday Institution (EP/S003053/1, grant number FIRG003).

**Competing Interests:** D.H. is co-founder of Brill Power Ltd., and is a technical advisor at Habitat Energy Ltd. V V. is a Technical Consultant at QuantumScape Corporation and Chief Scientist at Aionics Inc.

**Supplementary Information**

**Examples of the impact of successful open data ecosystems**

Openly available, high-quality data and associated software and community practices have significantly advanced research progress in many domains[9,10,12].

Open software libraries such as the Atomic Simulation Environment[86] and Pymatgen[87] provide broadly-accessible, well-tested tools to the materials science community for (i) standardization and running of workflows for numerous electronic structure codes,[88] and (ii) easy manipulation, analysis and sharing of input and output data. Such software libraries have in turn become foundational and spurred many data-centric innovations for computer-aided materials design,[15] from creation of large high-throughput computational databases[89] to adoption of atomistic machine learning.[15–18]

Drug design enjoys invaluable contributions stemming from advanced data sharing practices in protein crystallography. This community adopted both a common data format, mmCIF,[13] as a basis for public databases such as the Protein Data Bank (PDB), and common software.[9] In 1990, the International Union of Crystallography adopted the CIF format for data exchange.[14] By 1994, CIFs were required for all Acta Crystallographic C submissions; the Acta journals provided software packages to authors in order to assist file preparation in the desired format. Today, many journals require crystallographic data to be available in the PDB. *As a standard, CIFs revolutionized the ability to communicate key structural properties of solids.*

Another great triumph of open data is the human genome project. In the 1990s the commercial sector pursued proprietary genomic data for its potentially massive financial value, while parallel efforts were taking place in academia and research institutes, supported by National Institutes of Health (NIH) and other public agencies. The two sectors seemed to be competitors until Eric Lander proposed open access to the entire genome as the path to breakthroughs in biological sciences. Health care and the NIH and other public agencies supported his proposal and open-source data soon became available. The private sector learned that their proprietary discovery tools could create significant value when used in combination with the open genomic database – greater than that which would develop from a limited private database. Early successes spurred continued refinements in DNA sequencing, accelerating the rate thousandfold and reducing costs by a billionfold. Life science predictive tools based in this foundation have impacted understanding spanning from early-onset dementia to Covid-19.

The therapeutics field with complex heterogeneous data has a similar approach to our proposal to build a community based on open data and software.[90]

**Criteria for the impact of the Battery Data Genome**

It is easy to claim success in an endeavor in the absence of quantitative metrics. To avoid this, we propose BDG areas that can be tracked with the objective of estimating where successes have occurred. These metrics are not fixed nor meant to be all-inclusive but instead demonstrate the concept and start a conversation that may ultimately be refined in a workshop. Tracking these metrics addresses the need to quantify the impact of the expansive BDG vision and potentially also to identify mid-project issues that could benefit from adjustments in approach.

Success would be a rapid increase in each metric over the three phases of the BDG. Key metrics to track as the BDG is implemented might include:

- Number of interconnected and interoperative datahubs as part of the BDG
- Number of data sets uploaded to each datahub in the BDG

- Number of sharing institutions and contact individuals
- Total amount of data shared, and annual numbers on the extent of data downloaded from the BDG
- Use patterns including where and what types of data are viewed, downloaded and used and what type of organizations download data including industry, academia and other laboratories/ government organizations
- Publications which reference the BDG
- Publications which use BDG data
- Adoption of data and testing standards

In addition to these transactional numbers that can be pulled from publications and data hubs there are other more complex innovation metrics that might align with BDG success. These include the advancement of technology and increased adoption. While difficult to directly link to the BDG, they will help solidify need. Additional data on utility and benefit could also be captured using periodic community wide questionnaires.

**Cell and system metadata**

In order to enable rapid searching and filtering of data, and to enable consistency of comparisons between data, we suggest at least that the following minimum set of primary metadata relevant for all situations should be included when datasets are shared. (1) A globally unique identifier (UID) for each cell is required to enable cross referencing between databases, and to cell information. (2) To enable rapid data querying and filtering, we need to specify the objective of a test (e.g., calendar aging, fast charging optimization, EV drive cycles, square wave cycling), the cell chemistry of both electrodes, and nominal capacity. (3) Identification of test equipment used should be included to enable consistent comparisons. (4) Finally, contact details of the organization which published the data should be included. Standard formatting of primary metadata should be enforced to ensure interoperability. Table S1 shows an example of a metadata checklist that should be filled out when publishing a dataset.

**Table S1: Metadata Checklist**

| Primary Metadata | Example |
|---|---|
| Unique Identifier (UID) | 6 |
| Objective of Test | Calendar Aging |
| Anode Chemistry | Graphite |
| Cathode Chemistry | NMC |
| Nominal Capacity (Ah) | 2.0 |
| Test Equipment Used | Maccor 123 |
| Contact Details | abc@def.edu |

Rather than dictating a rigid structure, we instead suggest a number of areas that could be further developed by the community to become secondary metadata 'lego bricks', that can be slotted in as required:

- Cell design: Composition, additives, mass loading of each electrode
- Experiment design: Test protocols and techniques
- Constituent materials: Electrodes, electrolyte, binder, separator
- Geometric data: Electrode thicknesses, porosity, particle size and distribution
- Commercial cell data: Manufacturer, model name and number, link to datasheet
- Test setup details: Mechanical fixings, applied pressure, etc.
- References: Links to papers, reports, designs of experiments, protocols

Figure S1 shows an example of linked tables to describe constituent materials of a cell. This structure follows a JSON schema format as described in https://github.com/materials-data-facility/battery-data-schemas.

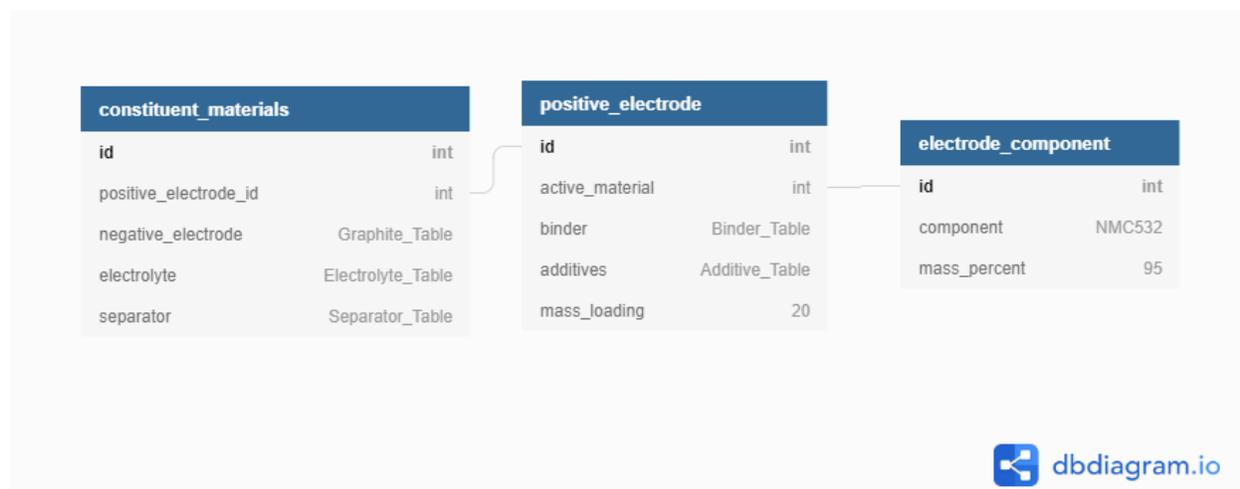

We have implemented a JSON metadata schema format on GitHub (included in the SI) to demonstrate this approach. To use this schema, each battery cell in a data hub would have an accompanying JSON metadata file. A list of primary metadata examples, a subsection of schema relationships, and the GitHub repository information can be found in the SI on cell and system metadata.

**Details of significant gaps in current battery data processing and analysis software**

Major software challenges in battery testing can be segmented into three stages: 1) protocol design and normalization 2) data management, and 3) post-processing and analysis.

1. *Protocol design and normalization:* Many testing procedures require standardization; here, we demonstrate principles by using battery life prediction as an example, since it is a gap that impacts the entire ecosystem. Building efficient and translatable battery cycling protocols (hardware-specific files that control cycling experiments) that are representative of a particular use-scenario is a near-universal challenge. Application spaces such as transportation offer detailed cell testing protocols while in other cases the protocols are undefined and ad-hoc programs are written by individual researchers. A community-supported software library that provides a framework for defining, reading and writing cycling protocols would a) accelerate the tailoring of testing to specific battery use modalities, b) create opportunities for cooperation whereby battery testing is shared across laboratories.[84]

2. *Data management:* Effective data stream management will be key to successful broad use of the BDG data hub, especially by the ML/AI community.[51,52,91,92] Growing from today's idiosyncratic workflows to community-maintained libraries for parsing, validating, structuring, and sharing data will require broad community support. The first objective of data management is to remove measurement anomalies, which requires significant software development; data quality metrics will need to be established. Software should next be able to convert "cleaned data" into the standardized formats. Finally software should ingest the data into the chosen datahub.

3. *Analysis using battery models (physics-based and data-driven):* We use cycle life prediction to illustrate this. With data cleaning completed in step (2) above, model building can typically focus on data transformation and feature engineering (e.g., the extraction of cycle-level summaries from large timeseries data[93,94]), and conversion into ML-ready formats -- all requiring significant software development. Modular libraries are needed for these new AI approaches that interface with existing data tools (e.g., the scientific Python stack[95]).

**Challenge problems driving community engagement and addressing broad issues – SI**

The first consideration in challenge problems is the topic area. As one familiar example we focus here on *cell level challenges*, which can be categorized into life, safety, performance, state of health, and manufacturing:

a) **Performance**: What metadata is most critical for successfully predicting device performance over a decade of cycling? Are algorithms for one cell chemistry extensible to others?

b) **State of Health**: Can capacity at any point in cell lifecycle be estimated from *brief* current/voltage excursions? What type of data and models best represent performance evolution during aging? What limits an algorithm's ability to predict future performance in any scenario and at any state of health?

c) **Life**: What are the minimum number of cycles needed to predict cycle life (for example to 80% of initial capacity) with different accuracies? How do cycle life predictions for one use-scenario transition to another – for example can fast charge data predict life for energy arbitrage on the grid? How well do predictions transfer from one chemistry to another – for example can Li-ion NMC cell data predict Iron Phosphate behaviors?

d) **Safety**: Is there a signal that indicates an impending thermal runaway? Can the thermal response of a cell during thermal runaway, as well as cell-to-cell variation, be predicted based on the properties of a cell (chemistry, format, energy- and power-density)? Can the combination of engineering models, performance data and ML provide insights into fault tolerant materials, components and cell designs?

e) **Materials to Manufacturing**: What are root causes of cell variability vs. performance metrics such as cycle life and temperature resistance? Can manufacturers solve inverse problems to optimize performance by improving cell design (e.g., electrode geometry, tab placement, etc.)?

Table S2 lists of some eighteen different challenge categories for the community to expand upon and prioritize, corresponding to the five cell level areas. Each category contains multiple challenges intended to address open questions. Safety[96] and other categories will require broader community discussions to reach consensus on priority directions and need for supporting benchmark datasets.

**Table S2 – Open questions guiding BDG Tier 2 cell challenge problems and experiments**

| Challenge Category | | Open Challenges |
|---|---|---|
| **Performance (P)** | **(P.1) Reference model** | (P.1.1) Compare the **accuracy and convenience** of data-driven, physics-based and reduced-order models. <br><br> (P.1.2) **Reproduce measurements** (e.g., current, voltage, temperature, mechanical and other non-electrochemical signals). <br><br> (P.1.3) Predict the available **energy and power**. |
| | **(P.2) Inhomogeneity** | (P.2.1) Identify **root causes of cell-to-cell variability** at beginning of life. <br><br> (P.2.2) **Predict temperature and electrical inhomogeneities**. <br><br> (P.2.3) **Design/control large-format** 3D-cell geometries and **multi-cell battery packs** with lessons from P.2.1 and P.2.2. <br><br> (P.2.4) Quantify **stochastic inhomogeneities** due to e.g., material, electrode, and manufacturing nonuniformities or flaws. |
| | **(P.3) Extensibility** | (P.3.1) Predict performance for each reference model type **for new chemistries and operating conditions** (duty cycles and temperature extremes). <br><br> (P.3.2) Demonstrate predictions on models calibrated to **single cells, for multi-cell systems**. |
| | **(P.4) Inverse problem** | (P.4.1) Optimize **operating limits** to avoid min/max voltages, transport saturation/depletion, side reaction windows, and electrochemical-mechanical-coupled damage. <br><br> (P.4.2) Optimize **fast-charge policies to avoiding damage** at beginning |
| | | of life. |
| State of Health (H) | **(H.1) Adaptivity** | (H.1.1) Evaluate reference performance model **accuracy at different levels of aging**.<br>(H.1.2) **Optimize operation policies under aging** (e.g., fast-charge or grid-dispatch policies) |
| | **(H.2) Diagnostics** | (H.2.1) **Classify and quantify degradation mechanisms** by tracking model parameter changes with aging (e.g. lithium inventory, active material, transport, kinetic, and ohmic impedance changes).<br>(H.2.2) Non-invasively **estimate the battery's full capacity SOH** (state-of-health) from real-world partial cycling data.<br>(H.2.3) Establish requirements on **electrochemical measurements** for quantifying and tracking health.<br>(H.2.4) Establish requirements on **non-electrochemical measurements** (mechanical/strain/pressure, acoustic, volume/diameter) for quantifying and tracking health. |
| Life (L) | **(L.1) Reference model** | (L.1.1) Compare the accuracy and convenience of data-driven, physics-based and reduced-order models.<br>(L.1.2) **Reproduce measurements** (e.g., current, voltage, capacity, resistance, and other features)<br>(L.1.3) **Classify and quantify degradation mechanisms** and paths. |

|  | **(L.2) Rapid prediction** | (L.2.1) **Reduce life test times** from today's 9-12 months to just 1-2 months. |
|  |  | (L.2.2) Identify **features with the highest correlation with specific degradation mechanisms**. |
|  |  | (L.2.3) Identify **features** - or combinations thereof - best describe and **forecast lifetime**. |
|  | **(L.3) Prognostics** | **Extrapolate laboratory-tests to real-world** usage, |
|  |  | (L.3.1) …from **simple to complex cycling**. |
|  |  | (L.3.2) …from **static to dynamic** (e.g., variable, path-dependent) **aging**. |
|  | **(L.4) Inhomogeneity** | (L.4.1) **Identify causes of growth of cell-to-cell variability**. Compare thermal and electrical inhomogeneities vs. **stochastic** ones. |
|  |  | (L.4.2) **Incorporate** inhomogeneity-driven and stochastic-driven **cell-to-cell variability growth into systems-level prognostics**. |
|  | **(L.5) Extensibility** | (L.5.1) Evaluate extensibility of **methods to other chemistries, electrode designs, cell designs, and use cases**. |
|  |  | (L.5.2) Extrapolate **materials-level data to higher length-scale** electrode- and device-level performance |
|  |  | (L.5.3) Extend incremental-capacity **synthetic-data methods to higher-rates** |
|  | **(L.6) Inverse problem** | (L.6.1) **Co-optimize battery utilization** (e.g., duty-cycles, charge protocols) and **lifetime**. |
|  |  | (L.6.2) **Optimize for operating limits that adapt with SOH** so that a desired lifetime can be achieved. |
|  |  | (L.6.3) Employ life predictions to inform maintenance. |
| **Safety** | **(S.1) Fault detection** | (S.1.1) **Detect impending faults** that may lead to **thermal runaway**. |

|   |   | Evaluate sensing requirements and response time reasonableness. |
|---|---|---|
|   |   | (S.1.2) Compare the **differences between healthy and unhealthy cells to enhance the fault-detection signal**-to-noise ratio. |
|   | **(S.2) Testing & design** | (S.2.1) **Given that the risk of failure is extremely low,** replicate rare failure scenarios and rapidly characterize cell fault-tolerance with physics-based models, experimental tests, and ML algorithms. |
|   |   | (S.2.3) Employ results to **guarantee safe packs and systems**. |
| **Materials-to-Manufacturing (M)** | **(M.1) Materials** | (M.1.1) **Interface the Battery Data, Materials Genome** and similar communities to more rapidly discover new materials, uncover materials structure/property relationships, better describe transport and reactions at interfaces, and provide feedback in the form of materials requirements. |
|   | **(M.2) Scale-Up** | (M.2.1) **Accelerate electrode, cell, and pack design** and qualification using data-driven, physics-based methods. |
|   |   | (M.2.2) Across the design space, **link life projections and cost.** |
|   | **(M.3) Manufacturing** | (M.3.1) Optimize cell/pack manufacturing processes, quality control, and cell formation processes. |
|   | **(M.4) Deployment** | (M.4.1) Optimize **system design and deployment** leveraging data-driven technologies. Validate against **real-world application and benchmark data** from industry. |
|   |   | (M.4.3) Assess **unmet needs of system integrators and end users**. Leverage data and 3rd party services to help address those needs. |

Once the topics are chosen, the next consideration is whether to use existing datasets (Tier 1) or use new datasets that must be developed for this purpose (Tier 2). The remainder of this section describes examples of Tier 1 challenge problems and associated data sets and additionally a Tier 2 dataset that addresses *multiple* challenge problems.

**Proposed Tier 1 Challenge Problem and Benchmark Dataset**

**Early cycle life prediction**: We propose a preliminary, Tier 1 challenge problem addressing early prediction of battery lifetime using existing experimental[26] and synthetic[27] datasets (corresponding to L.2, Rapid prediction). The objective is to achieve the lowest mean absolute error in the prediction of the number of cycles to failure given limited preliminary cycling data. Failure is defined as the number of cycles until capacity drops below 85% of its nominal value, or 0.935 Ah from 1.1 Ah. This level is selected as most cells from the Severson et al. study degraded to 0.935 Ah or lower. The experimental data will be supplemented with synthetic data generated according to Dubarry and Beck based on the same commercial cells (A123 Systems, model APR18650M1A) employed in the Severson et al. study. This dataset extensively explores degradation paths, degradation modes, and severities of degradation that are expected to correspond to different C-rates, temperatures, and DoD ranges. For each degradation path a C/10 charge-discharge cycle is provided every 10 cycles up to 200 cycles and then every 200 cycles up to cycle 3000. This represents more than 8 million voltage vs. capacity curves. The training dataset for this challenge will be composed of a selection of synthetic cycling data alone. The test set will comprise the Severson et al. dataset and a portion of the synthetic dataset. Challengers may choose the number of initial cycles considered in the training stage (corresponding to 1, 10, 50, 100, or 200 initial cycles).

**Safety:** We propose a Tier 1 challenge problem on predicting the range of heat-output from cells during thermal runaway under thermal abuse and nail penetration conditions using experimental data from the Battery Failure Databank.[97] The objective is to predict the range of heat output values for a cell of a particular cylindrical geometry (e.g. 18650, 21700, D-cell) with specific properties (electrode chemistries, energy-density, power-density) under either nail or thermal abuse conditions. As of February 2021, the Battery Failure Databank contains data from over 250 thermal runaway tests that employed a Fractional Thermal Runaway Calorimeter (FTRC) and simultaneous high-speed radiography. The FTRC facilitated decoupling of heat generated within the body of the cell and heat ejected from the cell, as well as quantification of ejected and non-ejected mass during thermal runaway. The Battery Failure Databank consists of numerous repeat tests for different cell geometries (18650, 21700, D-cell) from several different commercial manufactures. Values for the energy-density and power-density of each cell are provided. The accompanying high-speed radiography videos show the rate of thermal runaway propagation within each cell, providing researchers with insight to connect internal dynamic phenomena with external temperature and mass measurements. Challengers may choose specific cell properties for testing, such as geometry, energy-density, power-density, and electrode chemistry.

**Tier 2 Benchmark Dataset**

We propose a comprehensive dataset capable of informing a large portion of the challenge problems identified in Table S2. Such a dataset should be developed for at least one chemistry. Future tests on other cell designs/chemistries could then be less extensive, encompassing a smaller subset of tests. Table S4 outlines a preliminary proposal for tests to be conducted on the first cell technology to capture its physical design, chemistry, electrochemical, aging, and thermal abuse behavior. Mainly intended at beginning of life, some tests may also be repeated at the end of life to diagnose/confirm degradation mechanisms.

**Table S3 Physicochemical characterization of cell for Tier 2 benchmark dataset.**

| Category | *Task/*Measurement |
|---|---|
| Teardown | *Cut open cells in glovebox. Extract electrodes and* |
| | *store submerged in appropriate solvent. Extract electrolyte via centrifuge.* |
| Physical measurements | Cell outer dimensions |
| | Cell packaging dimensions and materials |
| | Jellyroll dimensions; number of stacks/winds |
| | Electrode thickness |
| | Separator thickness |
| | Cell component mass & volume |
| Microscopy (e.g., FIB-SEM, X-ray CT) | Active material particle size & morphology |
| | Electrode porosity |
| | Separator porosity |
| Chemical (e.g., SEM-EDX, ICP) | Active material composition |
| | Electrolyte composition |
| Electrochemical | *Punch electrode disks; Scrape off electrode material from backside of disk; Weigh each and record mass; Build half cells and full cells in coin cell format. Measure:*<br>• Pseudo-Open circuit potential (OCP) at C/100 rate<br>• Galvanostatic intermittent titration test (GITT)<br>• Electrochemical impedance spectroscopy (EIS)<br>• Rate capability<br>• Electrolyte conductivity |

|  | Thermal | Calorimetry during thermal runaway |
|--|---------|------------------------------------|
|  |         | Surface temperature measurements during thermal runaway |

* Electrochemical performance tests also to be performed on full commercial cell.

Due to the path dependence of battery degradation[98], a representative dataset must also test different stress factors and their possible combined effects. Table S5 details 8 different cell usage parameters and suggested levels for a proposed design of experiments. Varying all parameters is possible in an optimized design of experiments but would likely require more than 150 unique experiments. If each experiment is repeated 5 times to capture cell-to-cell variations, the proposed study would total 750 experiments for a single chemistry and design. The manuscript proposes to "crowd source" this benchmark data collection effort, distributing the burden across multiple labs and funding entities.

**Table S4 – Proposed parameters and parameter levels for the Tier 2 dataset.**

| Name | Units | Type | No. Levels | L1 | L2 | L3 | L4 | L5 |
|------|-------|------|-----------|----|----|----|----|----|
| Calendar aging portion | % of time | Discrete | 4 | 0 | 33 | 66 | 100 | |
| Charge cut off | % SOC | Discrete | 4 | | 60 | 80 | 90 | 100 |
| Depth of discharge | % DOD | Discrete | 4 | | 50 | 66.6 | 73.4 | 100 |
| Charge |  | Discrete | 3 | | 1 | 2 | 3 | |

| intensity | | | | | | | | |
|---|---|---|---|---|---|---|---|---|
| Discharge intensity | | Discrete | 3 | | 1 | 2 | 3 | |
| Temperature | oC | Nominal | 5 | 10 | 25 | 35 | 45 | Seasons |
| Charge profile | | Nominal | 5 | CC | Steps | Pulsing (solar) | Pulsing Mixed (solar) | Mixed inverted |
| Discharge profile | | Nominal | 5 | CC | Pulsing (Driving) | Pulsing (FR) | Mixed | Mixed Inverted |

**Proposed activities to launch the BDG**

Activity 1: *Broad community engagement: Host workshops for data segmentation and software interoperability to ensure needs and best practices of all communities are part of the inaugural design.* In the first-phase, *data and software categories will be refined* and relationships and standards that dictate design of interoperability between existing codes will be identified. The workshops will produce publicly available mission statements and listings of codes associated with the project, with opportunities for comment.

After categories and standards are defined, the next workshops will drive broad community engagement - core sharing platforms to manage proprietary information are essential for broad public/private collaboration. Examples include anonymizing sources while maintaining crucial metadata and sharing directly with those having common interests.

*Activity 2: Boot-strapping initial data hub growth*: Advance the development and population of each of the data hub spanning fundamental to field data. Databases such as the Battery Archive are an excellent first step towards realizing the BDG benefits and getting broad support. We suggest an initial focus on existing, published data. Many of the authors of this publication commit to providing data sets in different databases as the hubs evolve and to revisiting already published work over the next phase.

*Activity 3: Adopt sharing principles*: Parallel to (1) and (2), ensure data sharing practices align with publication requirements including adopting Findability, Accessibility, Interoperability, and Reusability (FAIR) data principles while also maintaining safeguards for product and intellectual property protection. A collaboration will only be accepted if appropriate safeguards are defined and implemented. For example, a 6–12-month embargo period after publication could be allowed to enable research groups to finish full analysis and protect appropriate intellectual property.

Other practices can be readily implemented to ensure set expectations are adopted – similar to the recent bottom-up approach in the battery community to expect specifics of cell designs such as Li metal thickness and quantity of electrolyte to be included in publications claiming rechargeable Li metal battery performance.[62,63] This would complement the top-down approach by publishers and funding agencies who require proof of data submission as part of the publication and work planning process. Initially the BDG will manage the complexities of existing data sources and individualized needs, but over time may evolve into standard approaches and routine protocols that govern future activities and data generation, and which are derived from experiential community development.